\documentclass[printer]{aa}
\usepackage{amssymb}
\usepackage{graphicx}
\usepackage{lscape}
\usepackage{txfonts}
\begin{document}
   \title{The Ultraviolet/optical variability of steep-spectrum radio quasars: the change of accretion rate ?}

%   \subtitle{I. Overviewing the $\kappa$-mechanism}

   \author{Minfeng Gu
%          \inst{1}
          \and
          Shuang-Liang Li%\inst{2,3}%\fnmsep\thanks{Just to show the usage
%          of the elements in the author field}
          }

   \institute{Key Laboratory for Research in Galaxies and Cosmology, Shanghai Astronomical Observatory,
    Chinese Academy of Sciences, 80 Nandan Road, Shanghai 200030, China\\
              \email{gumf@shao.ac.cn}
         %\and
         %    Department of Physics, University of California, Santa Barbara, CA 93106, USA
%         \and National Astronomical Observatories/Yunnan Observatory,
%             Chinese Academy of Sciences, P.O. Box 110, 650011 Kunming, Yunnan, China
%         \and
%             Key Laboratory for the Structure and Evolution of Celestial Objects,
%             Chinese Academy of Sciences, P.O. Box 110, 650011 Kunming, Yunnan, China
%             \thanks{The university of heaven temporarily does not
%                     accept e-mails}
             }

%   \date{Received September 15, 1996; accepted March 16, 1997}

% \abstract{}{}{}{}{}
% 5 {} token are mandatory

\titlerunning{The change of accretion rate in SSRQs}
\authorrunning{M. F. Gu \& S. L. Li}

  \abstract
{The steep-spectrum radio quasars (SSRQs) are powerful radio
sources, with thermal emission from accretion disk and jet
nonthermal emission likely both contributing in the Ultraviolet
(UV)/optical luminosity, however the former may play a dominant
role. While the UV/optical variability of SSRQs has been poorly
studied, little is known on the mechanism of their variability.}
{We investigate the mechanism of the UV/optical variability of
SSRQs.}
{A sample of eighteen SSRQs has been established in SDSS Stripe 82
region in our previous works, in which the flux and spectral
variability have been studied. In this work, we construct the
flux-flux diagram using SDSS $u$ and $i$ multi-epoch data for these
eighteen SSRQs. The standard accretion disk model is used to fit the
flux-flux variations, in order to explore the variability
mechanism.}
{The model fit to flux-flux diagram are tuned with fixed black hole
mass and varying accretion rate. We found that the flux-flux diagram
of all our SSRQs can be qualitatively described by the standard
accretion disk model with change of the accretion rate. Although non-thermal 
jet power-law emission can also qualitatively reproduce the variability,  the reasonable accretion
rates and black hole masses required to fit the flux-flux variations suggest that a disk emission
with variable accretion rate is a plausible description of the data.}
   {}


   \maketitle
%
%________________________________________________________________

\section{Introduction}

Although radio loud Active galactic nuclei (AGNs) only represent
10\% of whole AGNs population (e.g. Kellermann et al. 1989), they
are important in AGNs study, for example, to study the jet
production, composition, collimation and even the feedback to the
environment. As two subsets of radio-loud AGNs, flat-spectrum radio
quasars (FSRQs) and steep-spectrum radio quasars (SSRQs) are
different in many respects. FSRQs together with BL Lac objects, so
called Blazars, are the most extreme class of active galactic nuclei
(AGNs), characterized by strong and rapid variability, high
polarization, and apparent superluminal motion. These extreme
properties are generally interpreted as a consequence of non-thermal
emission from a relativistic jet oriented close to the line of
sight, which can well explain the large core-dominance in FSRQs in
general. In contrast, the SSRQs are usually lobe-dominated radio
quasars, and the radio lobe emission dominates over the radio core
emission. Their jets are viewed at larger angles than blazars,
hence, the beaming effects of jets are not severe (see e.g. Liu et
al. 2006), and the jet emission is not expected to dominate at
optical bands (e.g. Gu \& Ai 2011a, b).

AGNs exhibit variability at almost all wavelengths (e.g. Wiita 
1996; Vanden Berk et al. 2004). Multiwavelength
studies of variations in the radiation emitted from AGNs have played
important roles in exploring the physical conditions near the center
of AGNs. There have been extensive studies of the optical
variability of blazars (e.g. Ghisellini et al. 1997; Fan et al.
1998; Massaro et al. 1998; Ghosh et al. 2000; Clements \& Carini
2001; Raiteri et al. 2001; Villata et al. 2002; Vagnetti et al.
2003; Wu et al 2005, 2007; Gu et al. 2006; Hu et al. 2006; Poon et
al. 2009; Rani et al. 2010; Gu \& Ai 2011a). While it is generally
accepted that the nonthermal emission from relativistic jet oriented
close to the line of sight dominates the optical continuum, the
situation seems more complicated in FSRQs, with the evidence of
thermal emission in FSRQs (e.g. Pian et al. 1999; Grandi \& Palumbo
2004; Raiteri et al. 2007; D'Ammando et al. 2011), and the
redder-when-brighter trend found in several FSRQs (e.g. Gu et al.
2006; Rani et al. 2010; Gu \& Ai 2011a; Wu et al. 2011).

Intrinsically, the variations in the brightnesses of AGNs are
generally caused by physical variations in the jet and accretion
disk. However, the contribution of each component to the observed
variability varies from source to source. The properties of
steep-spectrum radio quasars (SSRQs) are in-between those of both
FSRQs and radio-quiet quasars. Therefore, it is not clear whether
the jet nonthermal or accretion thermal emission is primarily
responsible for the variation in SSRQs, even the situation can be
more complicated if two components are mixed together with
comparable contribution. However, the optical and spectral
variations of SSRQs have been poorly studied, and few explanations
have been presented (e.g. Stalin et al. 2004, 2005; Gu \& Ai 2011b),
in comparison to many investigations in radio-quiet AGNs (e.g.
Stalin et al. 2004, 2005; Gupta \& Joshi 2005; Wilhite et al. 2005;
Ai et al. 2010) and FSRQs. The optical variability of SSRQs has been
recently investigated by Gu \& Ai (2011b) for a sample of eighteen
SSRQs assembled from SDSS Stripe 82 region. As shown in Gu \& Ai
(2011b), the Mg II line to continuum luminosity ratio and the
anti-correlation between the variability at r band and the Eddington
ratio of SSRQs are all similar to those of radio-quiet AGNs. These
results strongly argued that the thermal emission from accretion
disk might be the dominant one in the optical continuum, and it may
be responsible for the variability of SSRQs. In terms of the
long-term variability, the change in accretion rate is used to
explain the optical variations in radio quiet AGNs (e.g. Li \& Cao
2008). In this work, we investigate whether the change of accretion
rate in standard accretion disk model can well describe the
Ultraviolet/optical variability of SSRQs.

The layout of this paper is as follows: in Section 2, we describe
the source sample; the results of model fit to the flux variations
are outlined in Section 3; Section 4 includes the discussion; and in
the last section, we draw our conclusions. The cosmological
parameters $H_{\rm 0}=70\rm~ km~ s^{-1}~ Mpc^{-1}$, $\Omega_{\rm
m}=0.3$, and $\Omega_{\Lambda}=0.7$ are used throughout the paper,
and the spectral index $\alpha$ is defined as
$f_{\nu}\propto\nu^{-\alpha}$, where $f_{\nu}$ is the flux density
at frequency $\nu$.

%__________________________________________________________________

\section{Sample}

Our sample of eighteen SSRQs consists of five SSRQs in Gu \& Ai
(2011a) and thirteen SSRQs in Gu \& Ai (2011b). The initial quasar
sample was selected as those quasars in both the SDSS DR7 quasar
catalogue (Schneider et al. 2010) and Stripe 82 region. The SDSS 
DR7 quasar catalogue consists of 105,783
spectroscopically confirmed quasars with luminosities brighter than
$M_{i}=-22.0$, with at least one emission line having a full width
at half-maximum (FWHM) larger than 1000 $\rm km~ s^{-1}$ and highly
reliable redshifts. The sky coverage of the sample is about 9380
$\rm deg^2$ and the redshifts range from 0.065 to 5.46. The
five-band $(u,~ g,~ r,~ i,~ z)$ magnitudes have typical errors of
about 0.03 mag. The spectra cover the wavelength range from 3800
$\rm \AA$ to 9200 $\rm \AA$ with a resolution of $\simeq2000$ (see
Schneider et al. 2010, for details). The Stripe 82 region extending over
right ascension $\alpha = 20^{\rm h} - 4^{\rm h}$ and declination
$-1^{\circ}.25 < \delta < +1^{\circ}.25$, was repeatedly scanned
during the SDSS-I phase (2000 - 2005) under generally photometric
conditions. This region was also scanned repeatedly over the course of three
3-month campaigns in three successive years in 2005 - 2007, as part
of the SDSS Supernova Survey (SN survey). The data are well-calibrated (Lupton et al. 2002). 

We cross-correlate the initial quasar sample with the Faint Images of
the Radio Sky at Twenty centimeters (FIRST) 1.4-GHz radio catalogue
(Becker, White \& Helfand 1995), the Green Bank 6-cm (GB6) survey at
4.85 GHz radio catalogue (Gregory et al. 1996), and the
Parkes-MIT-NRAO (PMN) radio continuum survey at 4.85 GHz (Griffith
\& Wright 1993). The initial quasar sample was first
cross-correlated with the SDSS quasar positions and the FIRST
catalogue to within 2 arcsec (see e.g. Ivezi\'{c} et al. 2002; Lu et
al. 2007). The resulting sample of SDSS quasar positions was then
cross-correlated with both the GB6 and PMN equatorial catalogues to
within 1 arcmin (e.g. Kimball \& Ivezi\'{c} 2008). Owing to the
different spatial resolutions of FIRST, GB6, and PMN, single or
multiple FIRST counterparts were found to within 1 arcmin, although
only a single GB6 and/or PMN counterpart existed. The radio spectral
index $\alpha_{\rm r}$ was then calculated between the single or
integrated FIRST and/or NRAO VLA Sky Survey (NVSS) 1.4 GHz within 1
arcmin and either or both of the GB6 and PMN 4.85 GHz (see details
in Gu \& Ai 2011a,b). We define a quasar to be a SSRQ according to
its radio spectral index $\alpha_{\rm r}>0.5$. Finally, five SSRQs
with single FIRST counterpart to within 1 arcmin of the SDSS
positions were presented in Gu \& Ai (2011a), while thirteen SSRQs
with multiple FIRST counterparts were listed in Gu \& Ai (2011b). 
To have the radio spectral information, we decided to consider 
only the sources with a radio counterpart in both 1.4 and 4.85 GHz radio 
images. In spite of this, our final sample, although not complete, still 
includes a quite large number (18)  of SSRQs. 
%To construct our sample, we require the detections in both 
%1.4 and 4.85 GHz, so the sources with detections only in either one
%radio band are excluded although some of them may have steep radio spectrum. 
%For this reason, our sample of eighteen SSRQs is not complete.

The source sample is listed in Table \ref{table_source}, in which
the black hole mass and the Eddington ratio from Gu \& Ai (2011b)
are shown. In calculating the Eddington ratio, the bolometric luminosity 
is estimated as $L_{\rm Bol} = 10 L_{\rm BLR}$ (Netzer 1990), in which 
the broad line region (BLR) luminosity $L_{\rm BLR}$ is derived following Celotti et al. (1997) by
scaling the strong broad emission lines $\rm H\beta$, Mg II, and C IV to 
the quasar template spectrum of Francis et al. (1991), in which
$\rm Ly \alpha$ is used as a flux reference of 100.

The multi-epoch photometric observations in the Stripe 82 region
enable us to investigate the optical variability of the selected
quasars. As in Gu \& Ai (2011a, b), we directly used the
point-spread-function magnitudes in the Catalog Archive Server (CAS) Stripe82 database from
the photometric data obtained during the SDSS-I phase from data
release 7 (DR7; Abazajian et al. 2009) and the SN survey during 2005
- 2007. We selected the sources classified as point sources in all
observational runs, and only data with good measurements
(high-quality photometry) were selected, of which the $ugriz$
magnitude are required to be brighter than the magnitude limit.
Finally, the data taken in cloudy conditions were also excluded.

\section{Results}

Besides the conventional flux versus color relation, the flux-flux
relation has been applied to examine the color variability for
quasars (e.g. Sakata et al. 2011; Schmidt et al. 2012). In this
work, to obtain a wide wavelength coverage in investigating the
spectral variability, the $u$ and $i$ magnitude in the same night
are used to plot flux-flux diagram. The reason of using $i$
magnitude instead of the $z$ magnitude as the longest wavelength is
that the signal-to-noise ratio is much higher in the former than
that of the latter for most targets. The corresponding wavelength of
SDSS $u$ and $i$ waveband in the source rest frame are shown in
Table \ref{table_source}.

The Eddington ratio $L_{\rm Bol}/L_{\rm Edd}$ of all SSRQs is $\rm -2.0 < log~ \it L_{\rm
Bol}/L_{\rm Edd} \rm < 0$ except for three sources (SDSS
J213004.75-010244.4, SDSS J012517.14-001828.9 and SDSS J015832.51-004238.2, see Table \ref{table_source}), which is
indicative of a standard thin disk (Shakura \& Sunyaev 1973) being
present in SSRQs. The low $L_{\rm Bol}/L_{\rm Edd}$ in SDSS
J213004.75-010244.4, and high values in SDSS J012517.14-001828.9 and SDSS J015832.51-004238.2, 
make the standard thin disc hardly applicable. However, the estimation of Eddington ratio is 
subject to large uncertainties both in black hole mass and BLR luminosity (see section 4). The standard accretion disk model used in Li \&
Cao (2008) has successfully explained the correlation between the
optical-UV variability amplitude and black hole mass found in Wold
et al. (2007) and Wilhite et al. (2008). In this work, we directly
adopted the model of Li \& Cao (2008) except that the temperature
distribution (the Equation (1) in Li \& Cao 2008) was modified as
\begin{equation}
T_{\rm eff} (r)=\{\frac{3GM \dot{M}}{8\pi \sigma
r^3}[1-(\frac{r_{\rm in}}{r})^{1/2}]\}^{1/4}
\end{equation}
(Frank, King \& Raine 2002). The temperature distribution in
Li \& Cao (2008) represents the disk temperature on the equatorial 
plane. In contrast, the temperature distribution we used here (in Equation (1)) 
describes the radial dependence of the effective temperature derived from 
the emergent flux from an accretion disk, which is more appropriate for
our purpose. We calculated accretion-disk spectra
assuming a steady geometrically thin, optically thick accretion disk
(Shakura \& Sunyaev 1973). In this case the emitted flux is
independent of viscosity, and each element of the disk face radiates
roughly as a blackbody with a characteristic temperature that
depends only on the mass of the black hole, $M_{\rm BH}$, the
mass accretion rate, $\dot{M}$, and the radius of the innermost stable
orbit (Peterson 1997; Frank, King \& Raine 2002). We have adopted
the Schwarzschild geometry (non-rotating black hole), and for this
the innermost stable orbit is at $r_{\rm in} = 6r_{\rm g}$, where
$r_{\rm g}$ is the gravitational radius defined as $r_{\rm g}
=GM_{\rm BH}/c^2$, $G$ is the gravitational constant, and $c$ is the
speed of light. Unlike blazars, SSRQs are expected to have the 
jet pointed away from the observer. We have assumed that the disk is viewed
with an inclination angle of $20^\circ$, which is the mean viewing angle of 
lobe-dominated radio quasars in Gu, Cao \& Jiang (2009). The accretion-disk spectrum is fully constrained by the two
quantities, the mass accretion rate and the black hole mass.

\begin{figure}
   \centering
   \includegraphics[width=1.\textwidth]{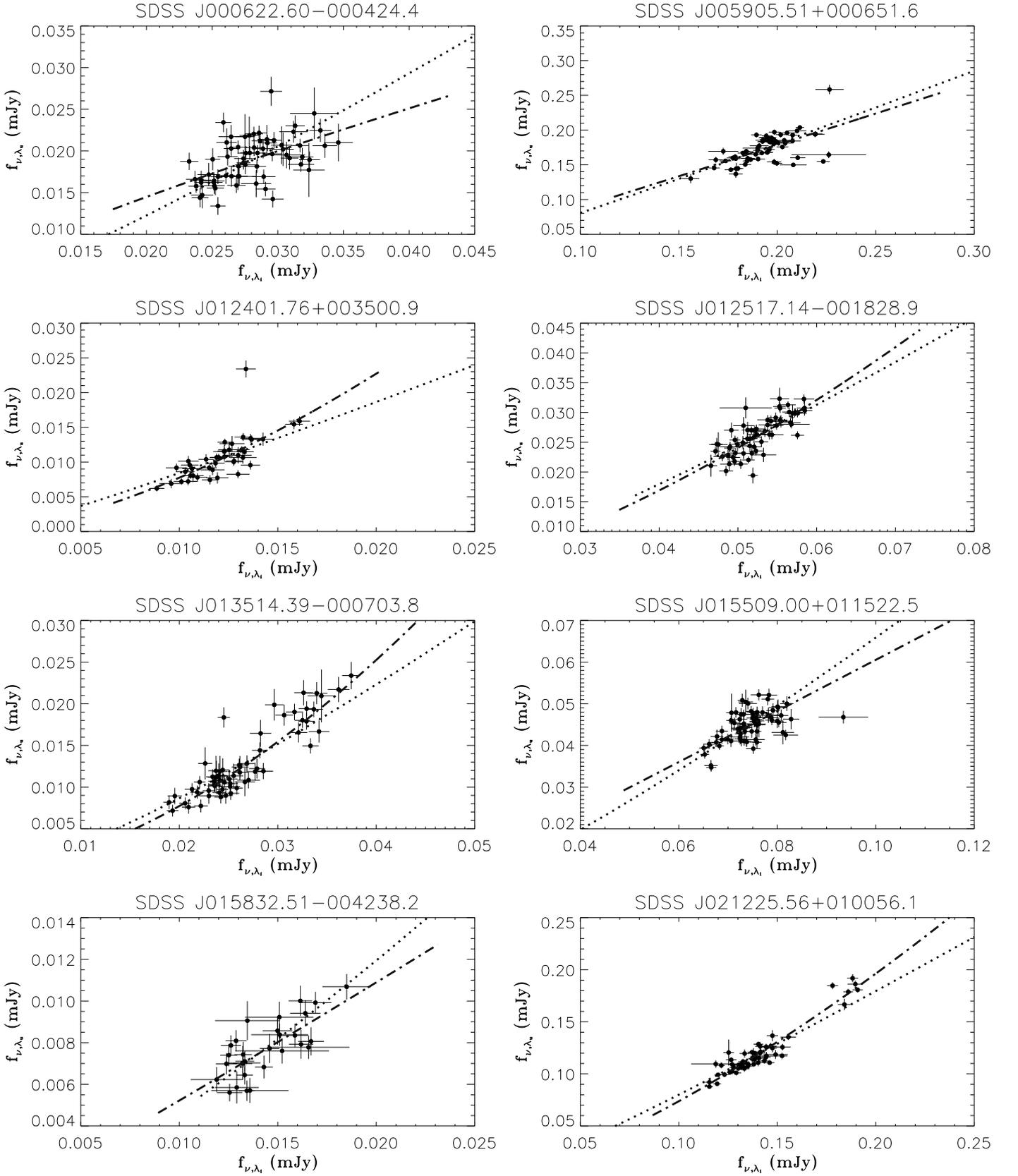}
   \caption{The flux-flux diagram for our SSRQs represented by the rest frame
   flux density at frequencies corresponding to $\lambda_u$
    and $\lambda_i$ (see Table \ref{table_source}) for $y$- and $x$-axis, respectively. The dotted line is
the best-fit standard accretion disk model with varied mass
accretion rates and a constant black hole mass, and the dash-dotted line represents the best-fit of 
a power-law. The solid circles
with error bars represent the multi-epoch flux data.}
              \label{ff1}%
    \end{figure}

We fit the accretion disk model to the flux-flux plots of individual
SSRQs with a free parameter of black hole mass and by changing the
mass accretion rate $\dot{M}$. The flux-flux diagrams are shown in Fig.
\ref{ff1} represented by the rest frame flux density at frequencies
corresponding to $\lambda_u$ and $\lambda_i$, with dotted lines
representing the best-fit accretion disk model. The best-fit black 
hole mass, reduced $\chi^2$ values and the range of the accretion rate $\dot{m}$ 
required to reproduce the flux-flux variations are listed in Table \ref{tbl1}. 
The accretion rate $\dot{m}$ is defined as $\dot{m}=\dot{M}/\dot{M}_{\rm Edd}$, in which $\dot{M}$
is the mass accretion rate from model fit, and $\dot{M}_{\rm Edd}$ is the Eddington accretion rate 
(the Eddington luminosity $L_{\rm Edd}=0.1\dot{M}_{\rm Edd}c^2$). Assuming $L_{\rm bol}=0.1\dot{M}c^2$, 
the accretion rate $\dot{m}$ is equal to the Eddington ratio $L_{\rm bol}/L_{\rm Edd}$. The 
standard thin disk seems to roughly describe the observed flux to flux variations. 
However, the $\chi^2$-test values are generally large. 
%(with the exception of two sources SDSS J015832.51$-$004238.2 and SDSS J024534.07$+$010813.7), 
%suggesting that the model parametrization of the spectral UV/optical emission may likely 
%be oversimplified. According to the $\chi^2$-test, 
%the accretion disk model is not rejected for only
%two sources (SDSS J015832.51$-$004238.2 and SDSS J024534.07$+$010813.7) at a 
%1\% level of significance. In these two sources, the UV/optical
%emission of SSRQs is likely from the thermal emission of accretion
%disk, and the UV/optical variability could be caused by the
%change of accretion rate.} 
%At first glance, the standard thin disk seems to give a good fit to the
%flux-flux data for all SSRQs by changing the mass accretion rate
%with a constant black hole mass.

As SSRQs are sources with a jet component, we also tested the possibility that a non-thermal 
power law emission can cause the observed flux-flux variations. As in the jet dominated blazars, 
the bluer-when-brighter trend is commonly observed, a power-law function of $f_{\rm u}=\alpha\times f_{\rm i}^{\beta}$ 
was adopted (see Sakata et al. 2011). The fitting of the power-law function to the flux-flux variations are presented in 
Fig. \ref{ff1}, and the best-fit parameters and the reduced $\chi^2$ values are listed in Table \ref{tbl1}. 
We found that the reduced $\chi^2$ value of the power-law fit is comparable to that of the standard 
thin disk for all quasars, although slightly improved. It seems difficult to
distinguish the thermal disk and nonthermal jet emission from a statistical point of view.

\section{Discussions}

While FSRQs are usually associated with core-dominated radio
quasars, SSRQs are generally related to lobe-dominated ones, usually
with two large-scale optically thin radio lobes. The beaming effect
is usually not strong in SSRQs because of the relatively large
viewing angle. As shown in Gu \& Ai (2011b), the Mg II line to
continuum luminosity ratio and the anti-correlation between the
variability at r band and the Eddington ratio of SSRQs are all
similar to those of radio quiet AGNs. These results strongly argued
that the thermal emission from accretion disk might be the dominant
one in the optical continuum, and it may be responsible for the
variability of SSRQs. While the change of accretion rate can qualitatively 
explain the optical variations in radio quiet AGNs (e.g. Li \& Cao
2008), it is expected that it can also be applied to SSRQs. From a qualitatively point of view, 
an accretion rate variations seems to work also for SSRQs. However, according to the $\chi^2$-test, 
only two sources are consistent with this expectation.

Quasars usually have strong broad emission lines (BELs), which could affect 
the photometry when they are in the SDSS $u$ and $i$ bands. Following
Elvis et al. (2012),  we found that only three strongest BELs have contribution larger than 
3\%, i. e. $\rm Ly\alpha$, $\rm H\alpha$ and $\rm H\beta$, for a typical bandwidth
of $1000~\AA$ for $u$ and $i$ bands (see equation (1) and Table 2 of Elvis et al. (2012)). 
Therefore, we only consider these three lines. In our sample, four sources (labelled in Table 1) are 
contaminated by BELs, of which $\rm Ly\alpha$ affects in two
sources (SDSS J012401.76$+$003500.9 and SDSS J213513.10$-$005243.8), $\rm H\alpha$
in SDSS J235156.12$-$010913.3, and $\rm H\beta$ in SDSS J021225.56$+$010056.1, 
respectively. In principle, the line contribution to the $u$ and $i$ fluxes could be  subtracted considering the SDSS 
spectra or the method proposed by Elvis et al. (2012). However, we prefer to not remove any line 
contribution. It is quite improbable that the large discrepancy observed between data 
and standard thin disc model can be entirely ascribed to emission line features. Note that a large reduced 
$\chi^2$ (6.4)  is also reported for SDSS J022508.07+ 001707.2, a SSRQ in the sample without  BELs contaminations.

%The large reduced $\chi^2$ value in other sources are probably due to the underestimation of the photometric errors. 
Since the reduced $\chi^2$ values are generally large, it is possible that the accretion disk model is oversimplified and  that variations of the accretion rate are not sufficient to explain the observed flux variability. As argued by Zuo et al. (2012), the change of accretion rate is important for producing the observed optical variability, but other physical mechanisms still need to be considered in modifying 
the simple accretion model for quasars. Schmidt et al. (2012) analyzed a large sample of 9093 
quasars from SDSS Stripe 82 region, and claimed that on timescales of years quasar
variability does not reflect changes in the mean accretion rate and
some other mechanism must be at work, presumably some disk
instability. The authors argued that this picture was confirmed by
the comparison of the observed color variability to sequences of
steady-state accretion disk models with varying accretion rates,
which exhibit much less color variability as a function of accretion
rate. While this is claimed for quasar ensembles, the thin accretion
disk models could well match the data both in color and in the
change of color with changing luminosity by changing the accretion
rate in some individual objects (see Fig. 11 in Schmidt et al. 2012). In contrast, Sakata et al. (2011) found that the
multi-epoch flux-to-flux plots could be fitted well with the
standard accretion disk model changing the mass accretion rate with
a constant black hole mass for luminous radio-quiet quasars. However, their 
sample size is rather small, only containing ten sources. Moreover, Pereyra
et al. (2006) found that the composite differential spectrum of two
epochs of observations for hundreds of SDSS QSOs in the rest-frame
wavelength range $1300-6000~\rm \AA$ can be successfully fitted by
the standard thin disk model, provided that their accretion rates
vary from one epoch to the next. It shoud be noted that the composite spectrum 
only represents the average characteristics of QSOs. It thus is unclear whether the scenario 
is applicable to the individual objects. Our results, although based on a smaller Stripe 82 region 
sample are in agreement with Schmidt et al. (2012).

As our sources are radio emitters, we can not exclude that the jet contributes to the optical/UV emission. 
The large reduced $\chi^2$ values then 
may naturally reflect the contribution of jet emission in the UV/optical band, in addition to
the thermal accretion disk emission. The nonthermal contribution to the MIR luminosity has
been investigated for a sample of powerful 3C RR galaxies and
quasars with the IRS and MIPS instruments on $Spitzer$ (Cleary et
al. 2007). Due to the low radio frequency selection, their quasars
are mostly steep spectrum sources, including several compact steep
spectrum sources (CSS). By fitting the continuum with synchrotron
(jet and lobe) and dust components, they found that nonthermal
processes can contribute a significant proportion (up to 90\%) of
the measured infrared emission in some quasars, although they also
shown that the nonthermal contribution may not be severe in the
sources with lower core dominance (see their Fig. 7), which is
typical in SSRQs. However, our disk model results (see Table 1) indicate that 
the accretion rates of the studied sources are in general agreement 
with values expected for standard accretion disks.

%We found from Table \ref{tbl1} that the accretion rate are all $\gtrsim0.01$, except for SDSS 
%J013514.39$-$000703.8, which have very weak $\rm H\beta$ but normal Mg II line. 
%It thus is reasonable to apply the standard thin disk to our sources. SDSS 
%J013514.39$-$000703.8 has the accretion rate lower than 0.01, therefore, is likely 
%subject to the radiation inefficient accretion flow rather than the standard thin disk. 
%Our disk model results (Table 1) indicate that the accretion rates of the studied sources are in general larger 0.01, as it is %expected for standard accretion disks. 
According to the Eddington 
ratio estimated using the bolometric luminosity from BLR, two sources %(SDSS J012517.14$-$001828.9 and SDSS J015832.51$-$004238.2) 
have Eddington ratio larger than $10^{0.5}\sim3$, and one source SDSS J213004.75$-$010244.4 has the Eddington ratio 
smaller than 0.01. On the contrary, the disk model fitted
accretion rate of these three sources are all well in the range of standard thin disk model $\gtrsim0.01$. 
The $\dot{m}$ difference is likely due to the large uncertainties in the black hole mass estimated from
BLR luminosity and the empirical relation. We estiamte the errors of the model fit black hole masses 
from the statistical uncertainty of $\chi^2$ fitting of the model, which is listed in Table \ref{tbl1}. While the error on the $M_{\rm
BH}$ estimates from BLR is $\sim0.4$ dex (e.g. Vestergaard \& Peterson 2006), the uncertainties of the black hole mass
from accretion disk model fit is much less. The black hole masses estimated from two
methods are compared in Fig. \ref{mbh12} for 17 SSRQs, after excluding the low accretion rate source SDSS J013514.39$-$000703.8. Generally, a good
consistence is found between these two masses. Therefore, in spite of the fact that the disk model fits
are not acceptable for most of the sources and a power law can also reproduce the observations, the reasonable
accretion rates and black hole masses required to fit the flux-flux
variations strongly suggest that a disk emission with variable
accretion rate is a plausible description of the data.

%The CSS are young radio sources, thought to be at the early stage of
%evolution of radio jets, with jets having no enough time to expand
%to larger scales (O'Dea 1998).  It has been argued that the
%optical-UV emission of the GPS/CSS sources is typical of broad-line
%quasars. The SED through radio to X-ray band shows that both broad
%emission lines and a big blue bump are present in CSS, and there is
%no signature of a jet synchrotron emission in the optical-UV band,
%although they are strong radio emitters (Siemiginowska et al. 2008). 
%The five SSRQs (SDSS J012401.76$+$003500.9, SDSS J012517.14$-$001828.9, 
%SDSS J015832.51$-$004238.2, SDSS J021728.62$-$005227.2, and SDSS J022508.07$+$001707.2) 
%in Gu \& Ai (2011a) may be good candidates of CSS, 
%due to the fact that only a single counterpart is found in FIRST image 
%within 1 arcmin of SDSS positions (see also Gu \& Ai 2011b). However, we 
%found that the accretion disk model is acceptable in only one source, 
%i. e., SDSS J015832.51$-$004238.2.

Optical polarimetry could be extremely useful in investigating 
the nature of optical/UV emission in radio sources. 
As shown by the recent studies of 3C 120 and 3C 111, the low optical
polarization supports the thermal origin of the optical emission
(Chatterjee et al. 2009, 2011), however, the detection of
polarization percentages of 3\% or even higher in 3C 111 suggests
that during particular activity phases an additional contribution
from non-thermal emission from the jet may be present (Jorstad et
al. 2007; Chatterjee et al. 2011). Motivated by this fact, the
polarimetry observation are therefore needed to further explore the
jet contribution in the optical band for SSRQs, in addition to the
photometric and spectroscopic monitoring.

%It is believed that the timescale of changes of the accretion rate
%is about the viscous timescale (e.g. Pringle 1981; Frank et al.
%2002), which is about $10^6$ years for our quasars, and much longer
%than the timescale of the flux variation we observed. Therefore, it
%is rather difficult for the change of accretion rate to be a primary
%solution for the variations. Alternatively, the variations of the
%characteristic temperature of an accretion disk can be responsible
%for the flux variations. Gaskell (2008) use the optical variability
%time-scales of quasars to argue that the variations must propagate
%at close to the speed of light, rather than on viscous time-scales.

\section{Summary}

We explore the flux-flux diagram using SDSS $u$ and $i$ multi-epoch
data for a sample of eighteen SSRQs in SDSS Stripe 82 region
established in our previous works. The standard accretion disk model
is used to fit the flux-flux variations, in order to explore the
variability mechanism. We found that the flux-flux diagram
of all our SSRQs can be qualitatively described by the standard
accretion disk model with change of the accretion rate. Although non-thermal 
jet power-law emission can also qualitatively reproduce the variability,  the reasonable accretion
rates and black hole masses required to fit the flux-flux variations suggest that a disk emission
with variable accretion rate is a plausible description of the data.

%                                                One column figure
%----------------------------------------------------------- S_vib
%   \begin{figure}
%   \centering
%   \includegraphics[width=\textwidth]{mgcon.eps}
%      \caption{              }
%         \label{mgcon}
%   \end{figure}
%
%______________________________________________________________

\begin{acknowledgements}
We thank the referee for constructive comments that greatly
improved the manuscript.
MFG thanks Y. Ai for the help on data analysis, and S. Shen for helpful discussions. This work is
supported by the 973 Program (No.
2009CB824800), and by the National Science Foundation of China (grants
10833002, 10903021, 11073039 and 11233006). Funding for the SDSS and SDSS-II was provided by the
Alfred P. Sloan Foundation, the Participating Institutions, the
National Science Foundation, the U.S. Department of Energy, the
National Aeronautics and Space Administration, the Japanese
Monbukagakusho, the Max Planck Society, and the Higher Education
Funding Council for England. The SDSS Web site is
http://www.sdss.org/.
\end{acknowledgements}

\clearpage

\newpage

\begin{landscape}
\begin{table}
\caption{\label{table_source}Source list: Col. 1 - SDSS source name;
Col. 2 - redshift; Cols. 3 - 4 - the corresponding wavelength of SDSS $u$ and $i$
wavebands in the source rest frame, respectively ($^{e,~ f, ~g}$ the prominent broad emission lines 
$\rm Ly\alpha$, $\rm H\beta$, and $\rm H\alpha$ in the photometric wavebands, respectively); Col. 5 -
$^a$ black hole mass in unit of solar mass estimated from the empirical relation of
BLR radius and the BLR line width; Col. 6 - the Eddington ratio
$l=L_{\rm BOL}/L_{\rm EDD}$ with the black hole mass in Col. 5, and the bolometric 
luminosity estimated from BLR luminosity; Cols. 7 - 8 - $^{b}$ black hole mass in unit of solar mass, and the range of
the accretion rate from the model fit with a standard accretion disk model, respectively; Col. 9 - $^c$ the reduced $\chi^2$ of the accretion disk model fit; 
Col. 10 - $^d$ the reduced $\chi^2$ of the power-law fit; Cols. 11 - 12 - the parameters of the power-law fit $f_{\rm u}=\alpha\times f_{\rm i}^{\beta}$.} \centering
\begin{tabular}{lcclcccccccc}
\hline\hline
SDSS source   &    $z$   & $\rm \lambda_{u}$ & $\rm \lambda_{i}$ & log $(M_{\rm BH})^a$ & log $(l)$ & log $(M_{\rm BH})^b$ & log $(\dot{m})$ & $(\chi^2/\nu)^c$ & $(\chi^2/\nu)^d$ & $\alpha$ & $\beta$ \\
 & & ($\AA$) & ($\AA$) & ($\rm M_{\odot}$) & & ($\rm M_{\odot}$) & & & & (mJy) & \\ %L_{\rm BOL}/L_{\rm EDD}
(1)&(2)&(3)&(4)&(5)&(6)&(7)&(8)&(9)&(10)&(11)&(12)\\
\hline
J000622.60$-$000424.4  &   1.0377  &  1742.65      &  3671.30      &  8.92   &   -1.16  &  8.85$^{+0.09}_{-0.10}$ & [-1.68,-1.45]  & 3.05  &  2.65  &  0.323$^{+0.009}_{-0.010}$  &  0.794$^{+0.008}_{-0.007}$  \\
J005905.51$+$000651.6  &   0.7189  &  2065.86      &  4352.20      &  8.96   &   -1.15  &  8.86$^{+0.08}_{-0.06}$ & [-1.21,-0.97]  & 8.23  &  8.06  &  0.910$^{+0.012}_{-0.009}$  &  1.011$^{+0.006}_{-0.008}$  \\
J012401.76$+$003500.9  &   1.8516  &  1245.27$^e$  &  2623.44      &  9.36   &   -1.36  &  8.49$^{+0.16}_{-0.18}$ & [-0.83,-0.51]  & 4.80  &  4.37  &  9.876$^{+0.336}_{-0.336}$  &  1.553$^{+0.008}_{-0.006}$  \\
J012517.14$-$001828.9  &   2.2780  &  1083.28      &  2282.18      &  8.50   &    0.52  &  9.31$^{+0.05}_{-0.04}$ & [-0.92,-0.76]  & 2.19  &  2.11  &  2.760$^{+0.055}_{-0.055}$  &  1.583$^{+0.006}_{-0.006}$  \\
J013514.39$-$000703.8  &   0.6712  &  2124.82      &  4476.42      &  8.41   &   -1.60  &  8.98$^{+0.10}_{-0.10}$ & [-2.52,-2.21]  & 2.03  &  1.72  &  6.187$^{+0.260}_{-0.266}$  &  1.709$^{+0.012}_{-0.010}$  \\
J015509.00$+$011522.5  &   1.5480  &  1393.64      &  2936.03      &  8.64   &   -0.20  &  9.21$^{+0.05}_{-0.05}$ & [-1.14,-0.91]  & 2.42  &  2.24  &  0.629$^{+0.011}_{-0.011}$  &  1.017$^{+0.006}_{-0.006}$  \\
J015832.51$-$004238.2  &   2.6071  &   984.45      &  2073.97      &  8.29   &    0.50  &  8.99$^{+0.13}_{-0.13}$ & [-1.00,-0.76]  & 1.00  &  0.93  &  0.691$^{+0.043}_{-0.041}$  &  1.061$^{+0.014}_{-0.014}$  \\
J021225.56$+$010056.1  &   0.5128  &  2347.30      &  4945.13$^f$  &  8.77   &   -1.37  &  8.78$^{+0.07}_{-0.07}$ & [-1.68,-1.37]  & 4.00  &  2.81  &  1.904$^{+0.023}_{-0.029}$  &  1.412$^{+0.007}_{-0.006}$  \\
J021728.62$-$005227.2  &   2.4621  &  1025.68      &  2160.83      &  8.84   &   -0.55  &  9.09$^{+0.10}_{-0.10}$ & [-1.29,-1.05]  & 1.89  &  1.87  &  1.376$^{+0.063}_{-0.061}$  &  1.267$^{+0.010}_{-0.010}$  \\
J022508.07$+$001707.2  &   0.5270  &  2325.47      &  4899.15      &  8.88   &   -1.80  &  9.03$^{+0.04}_{-0.05}$ & [-2.13,-1.97]  & 6.40  &  5.95  &  1.995$^{+0.038}_{-0.040}$  &  1.525$^{+0.009}_{-0.008}$  \\
J023313.81$-$001215.4  &   0.8072  &  1964.92      &  4139.55      &  8.41   &   -0.95  &  8.77$^{+0.08}_{-0.10}$ & [-1.83,-1.59]  & 3.05  &  2.68  &  4.997$^{+0.115}_{-0.110}$  &  1.573$^{+0.006}_{-0.006}$  \\
J024534.07$+$010813.7  &   1.5363  &  1400.07      &  2949.57      &  9.65   &   -1.81  &  9.68$^{+0.07}_{-0.06}$ & [-1.83,-1.67]  & 0.63  &  0.56  &  0.189$^{+0.008}_{-0.008}$  &  0.954$^{+0.015}_{-0.015}$  \\
J213004.75$-$010244.4  &   0.7040  &  2083.92      &  4390.26      & 10.08   &   -2.35  &  9.43$^{+0.04}_{-0.04}$ & [-2.05,-1.89]  & 2.98  &  2.91  &  1.368$^{+0.022}_{-0.021}$  &  1.652$^{+0.008}_{-0.008}$  \\
J213513.10$-$005243.8  &   1.6548  &  1337.58$^e$  &  2817.91      &  8.58   &   -0.09  &  9.29$^{+0.03}_{-0.05}$ & [-0.91,-0.75]  & 4.81  &  4.51  &  2.914$^{+0.038}_{-0.038}$  &  1.701$^{+0.005}_{-0.005}$  \\
J221409.96$+$005227.0  &   0.9078  &  1861.31      &  3921.27      &  9.05   &   -1.31  &  8.67$^{+0.10}_{-0.10}$ & [-1.21,-0.97]  & 2.97  &  2.83  &  2.135$^{+0.041}_{-0.034}$  &  1.305$^{+0.005}_{-0.007}$  \\
J231607.25$+$010012.9  &   2.6291  &   978.48      &  2061.39      &  9.15   &   -0.42  &  9.28$^{+0.06}_{-0.04}$ & [-0.60,-0.45]  & 2.22  &  2.08  &  2.557$^{+0.049}_{-0.046}$  &  1.549$^{+0.006}_{-0.006}$  \\
J233624.04$+$000246.0  &   1.0949  &  1695.07      &  3571.05      &  9.16   &   -1.00  &  9.03$^{+0.06}_{-0.08}$ & [-1.21,-0.97]  & 4.65  &  2.09  &  3.099$^{+0.046}_{-0.046}$  &  1.617$^{+0.006}_{-0.006}$  \\
J235156.12$-$010913.3  &   0.1739  &  3024.96      &  6372.77$^g$  &  8.90   &   -1.53  &  9.03$^{+0.02}_{-0.05}$ & [-1.84,-1.60]  & 10.83 &  10.63 &  0.784$^{+0.006}_{-0.008}$  &  1.102$^{+0.008}_{-0.010}$  \\
\hline
\label{tbl1}
\end{tabular}
\end{table}
\end{landscape}

\clearpage

\newpage

\addtocounter{figure}{-1}

\clearpage

\newpage

\begin{figure}
   \centering
   \includegraphics[width=1.\textwidth]{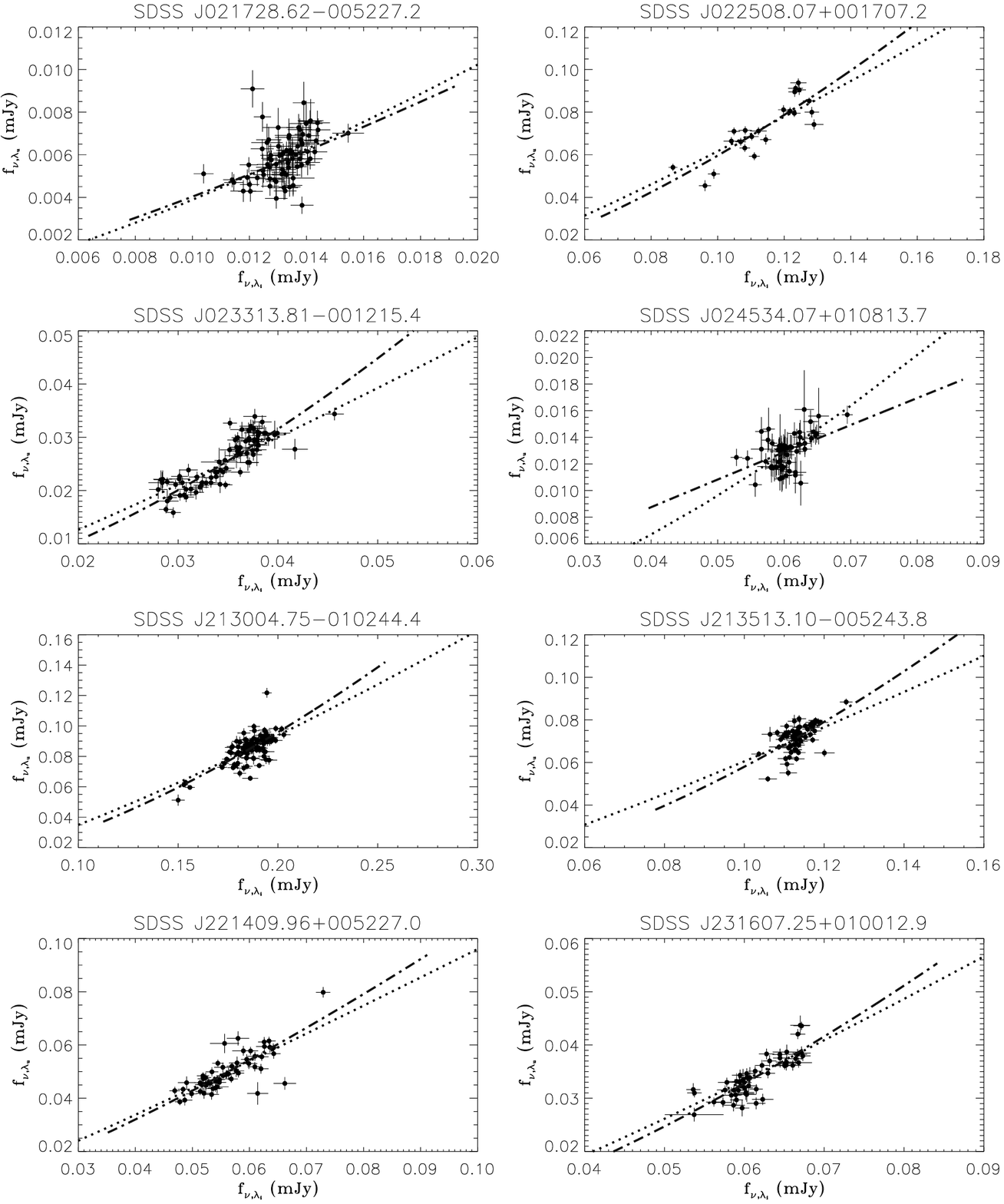}
   \caption{- $continued$. }
              \label{ff2}%
    \end{figure}

\addtocounter{figure}{-1}

\clearpage

\newpage

\begin{figure}
   \centering
   \includegraphics[width=1.\textwidth]{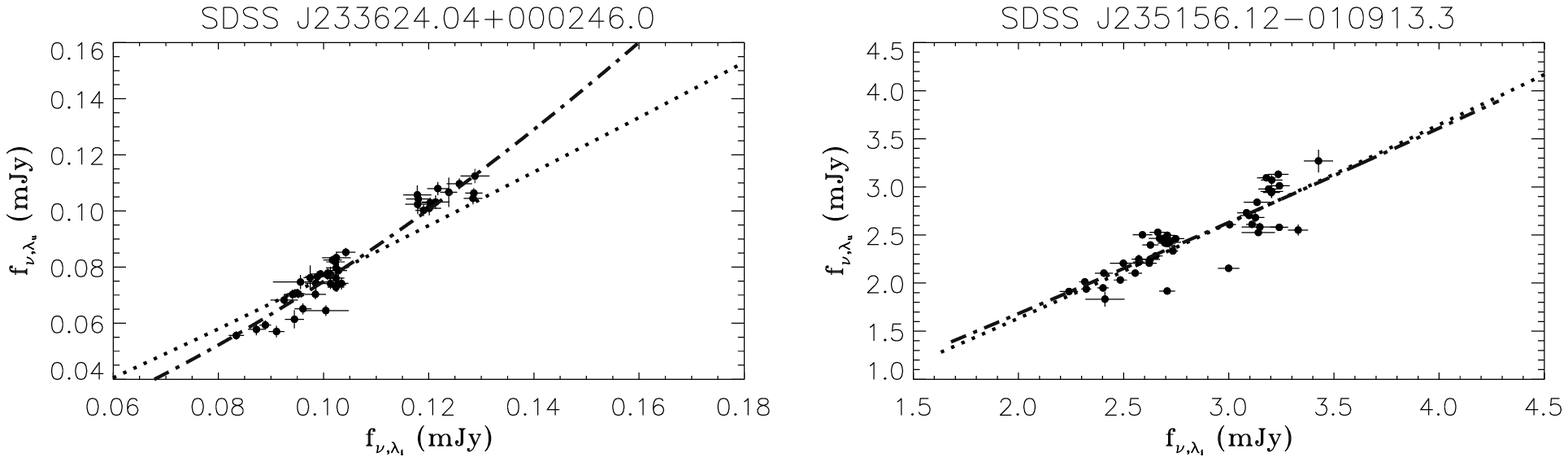}
   \caption{- $continued$. }
              \label{ff3}%
    \end{figure}

\begin{figure}
   \centering
   \includegraphics[width=.8\textwidth]{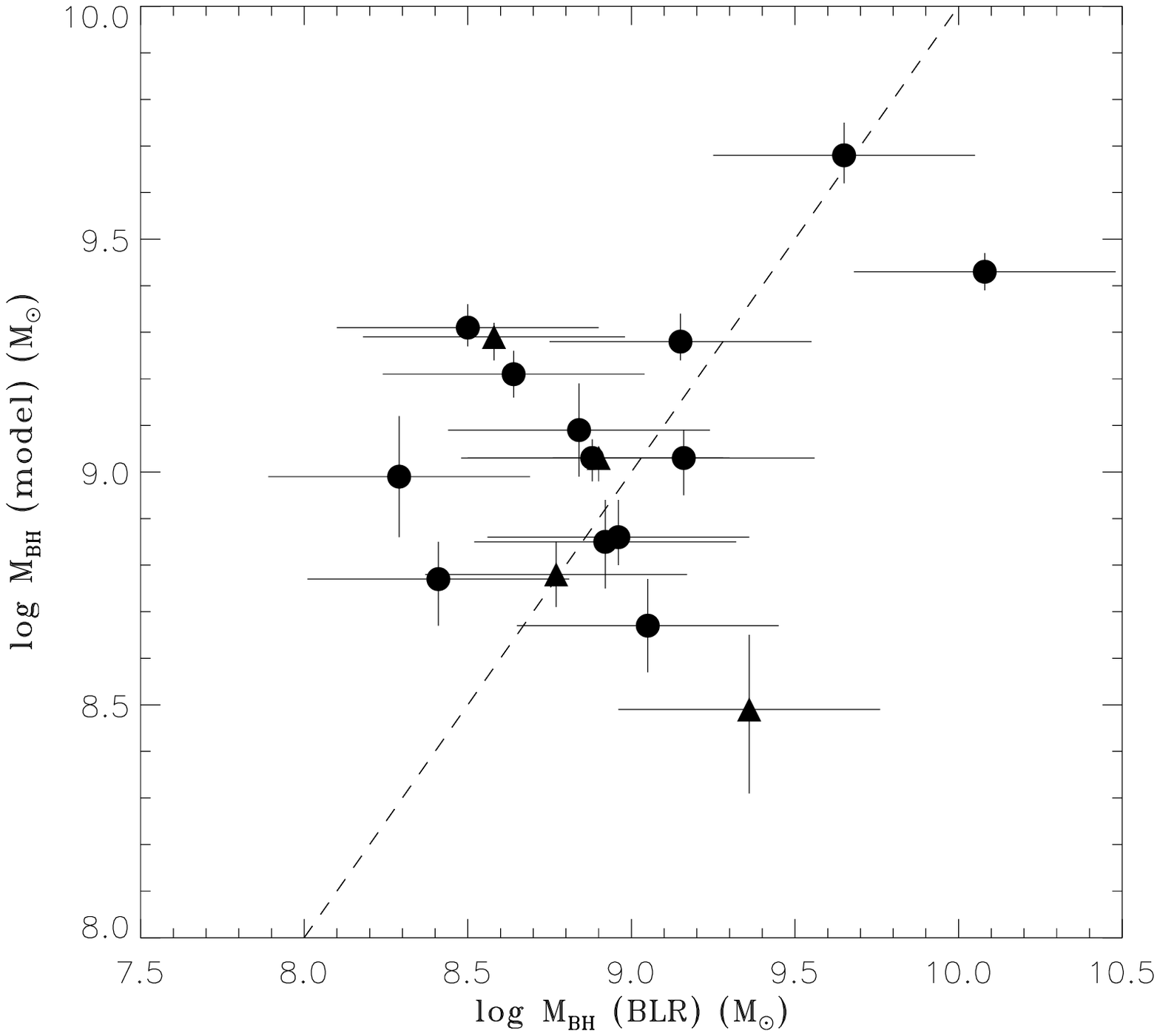}
   \caption{The black hole mass estimated from the empirical relation of BLR radius and the
BLR line width compared with that of the best-fit standard accretion disk model for 17 SSRQs 
after excluding the source with low accretion rate SDSS J013514.39$-$000703.8. The solid triangles are 
four sources with strong emission line contamination in either $u$ or $i$ bands.}
              \label{mbh12}%
    \end{figure}

\end{document}